# The Rotation Period of Comet 29P/Schwassmann–Wachmann 1 Determined from the Dust Structures (Jets) in the Coma

A. V. Ivanova[a], V. L. Afanasiev[b], P. P. Korsun[a], A. R. Baranskii[c], V. Andreev[a, d, e], and V. A. Ponomarenko[c]

[a] Main Astronomical Observatory, National Academy of Sciences of Ukraine, Goloseevo, Kiev, 03680 Ukraine
[b] Special Astrophysical Observatory RAS, Nizhni Arkhyz, Zelenchuk region, Karachai-Cherkessian Republic, 369167 Russia
[c] Astronomical Observatory, Kiev National University, Observatornaya ul. 3, Kiev, 04053 Ukraine
[d] International Center of Astronomical, Medical, and Ecological Research, ul. Akademika Zabolotnogo 27, Kiev, 03680 Ukraine
[e] Terskol Branch of Astronomy Institute RAS, Elbrus region, Karachai-Cherkessian Republic, 361603 Russia
R

**Abstract:** The results of the photometric observations of comet 29P/Schwassmann–Wachmann 1 are ana- lyzed. The comet demonstrates substantial activity at heliocentric distances larger than 5 AU, i.e., outside the water ice sublimation zone. The CCD images of the comet were obtained in wideband *R* filters at the 6-meter telescope of the Special Astrophysical Observatory of the Russian Academy of Sciences (SAO RAS) and at the 2-meter Zeiss-2000 telescope of the Peak Terskol Observatory. The processing of the images with special digital filters allowed the active structures (jets) to be distinguished in the dust coma of the comet. With the cross-correlation method, the rotation period of the cometary nucleus was determined as 12.1 ± 1.2 and 11.7 ± 1.5 days for the observations taken in December 2008, and February 2009, respectively. The probable causes of the difference in the estimates of the rotation period of the cometary nucleus obtained by different authors are discussed.

## INTRODUCTION

Comet 29P/Schwassmann–Wachmann 1 (SW1) was discovered by German astronomers A. Schwassmann and A. Wachmann in 1925. The semimajor axis of the orbit is 5.99 AU, the eccentricity is 0.045, and the orbital period of the comet is 14.66 years. The estimates of the radius of the comet show that its nucleus is larger than those of the typical short-period comets. The diameter of the comet SW1 nucleus, 15–44 km, is estimated from the analysis of the thermal radiation at 20 μm (Cruikshank Brown, 1983) and from the visible photometry (Meech et al., 1993).

Comet SW1 is referred to the Centaurus class, the members of which are thought to be "escapees" from the Kuiper belt. Over many years of observations, the comet episodically has demonstrated flare activity (Richter, 1941; 1954; Roemer, 1958; 1962; Whipple, 1980). Jewitt (1990) noted that the coma had never completely disappeared in spite of different degrees of comet activity during the whole period of observations. During the comet SW1 survey from 2002 to 2007, Trigo-Rodríguez et al. (2008) detected 28 flares from the comet. The authors note that the typical flares from the comet, when its brightness sharply increases by 1–4 magnitudes, occur 7.3 times per year. No signs of periodicity in the flare activity of the comet have been detected, and the causes of the nonperiodic character of the flares are still obscure. As for the estimates of the rotation period of comet SW1, the unambiguous result has not been obtained until now. The first attempts to determine the rotation period of the comet were made many years ago (Whipple, 1980; Meech et lal., 1993; Luu and Jewitt, 1993; Cabot et al., 1997). From the recent studies devoted to this subject, the paper by Stansberry et al. (2004) can be marked out. With the use of the structures observed in the cometary coma, Stansberry et al. (2004) estimated the rotation period of comet SW1 as 60 days. However, as the authors themselves reported, this estimate of the rotation period contradicts such earlier obtained values as 6 days (Jewitt, 1990) and 14 hours (Meech et al., 1993). A similar value for the period, 50 days, was derived by Spanish researchers from the above-mentioned data of two-year observations of the comet (Trigo-Rodríguez et al., 2010).

In the present study, to estimate the rotation period of the comet, and to compare it with the earlier results, we examine the active structures in the dust coma of comet SW1 seen in the images obtained at different times at the Large Alt-Azimuth Telescope (LAT; the mirror diameter is 6 m) of the Special Astrophysical Observatory of the Russian Academy of Sciences (SAO RAS) and at the 2-m telescope of the Peak Terskol Observatory of the International Centre of Astro-

Observation logbook

| Date, observation period, UT | Frame number | Δ, AU | r, AU | P[a] |
|---|---|---|---|---|
| Dec. 2008, 3.130–3.323 | 15 | 5.413–5.401 | 6.084–6.085 | 280.6–280.4 |
| Dec. 2008, 4.115–4.501 | 13 | 5.401–5.389 | 6.085 | 280.3 |
| Feb. 2009, 14.901–14.903 | 3 | 5.244 | 6.106 | 104.4 |
| Feb. 2009, 18.863–18.865 | 3 | 5.272 | 6.107 | 103.6 |
| Feb. 2009, 19.721–19.725 | 5 | 5.292 | 6.107 | 103.4 |

[a] The position angle of the extended radius-vector is shown in degrees.

nomical, Medical, and Ecological Research (IC AMER).

## OBSERVATIONS AND PROCESSING

On 3–4 December, 2008, comet SW1 was observed with the SCORPIO focal reducer in the prime focus ($F/4$) of the 6-meter telescope (LAT SAO RAS) (Afanasiev and Moiseev, 2005). The SCORPIO focal reducer was used in the photometric mode. The CCD camera with a matrix of $2048 \times 2048$ pixels was a receiver. The receiver's field of view was $6.1' \times 6.1'$, and the image scale was $0.18''$ per pixel. The comet was observed during a period of high activity with the use of a wideband $R$ filter. Twenty eight images of the comet were obtained with a 60-s exposure. To increase the signal-to-noise ratio, $2 \times 2$ neighbouring pixels were grouped together in bins (the binning procedure).

On 14–19 February, 2009, comet SW1 was observed with the focal reducer mounted in the Cassegrain focus ($F/8$) of the 2-meter Zeiss 2000 telescope of the Peak Terskol Observatory (IC AMER). The CCD Photometrics camera equipped with a matrix of $562 \times 562$ pixels cooled with liquid nitrogen was a receiver. Its field of view was $8.5' \times 8.5'$, and the image scale was $0.99''$ per pixel. The images of the comet were obtained with a wideband $R$ filter.

The acquired data were reduced with the IDL codes (http://www.ittvis.com/idl). The preprocessing of the observational data included the accounting for the matrix bias, the cleaning of the images from the traces of cosmic particles, and accounting for the flat fields. The flat fields were obtained from the images of the morning sky.

Twenty eight images of comet SW1 were obtained at the 6-meter telescope during three nights in December 2008. Eleven images obtained at the 2-meter telescope in February 2009 allow us to follow the change in the coma structure during the period of five days.

The information about the observations (the observation time, the number of frames, the helio- and geocentric distances, and the position angle) is presented in detail in Table 1 and Fig. 1.

## SELECTION OF JETS IN THE COMA IMAGES

To select the weak-contrast structures (jets) in the images of the dust coma of the comet, we used a special code called Astroart (http://www.msb-astroart.com/), which is provided with a number of digital filters: the unsharping mask, the Gaussian blur, and the Larson–Sekanina filter (Larson and Sekanina, 1984).

**The unsharp mask.** Its use increases the contrast at the brightness edges. The processing technique with this filter is based on superimposing the original image and its blurred inverted copy.

**The Gaussian blur.** The filter is based on the Gaussian distribution of the surface brightness; its formula (for two dimensions) is the following

$$G(u, v) = \frac{1}{2\pi\sigma^2} e^{-(u^2+v^2)/(2\sigma^2)},$$

where $r^2 = u^2 + v^2$ is the radius of blurring and $\sigma$ is the Gaussian distribution deviation.

The equation describes the concentric circles distributed according to the Gaussian formula with respect to the image centre. The value of each of the pixels is averaged relative to the neighboring pixels as a function of distance. The above-mentioned filters were used in the following way. First, the original image was processed with the unsharp mask with a radius of 1.5 pixels. Then, a copy of the image obtained was processed with the Gaussian blur with a radius of 0.7 pixel. The result was the ratio of the first image to that processed with the Gaussian blur was obtained.

**The Larson–Sekanina filter.** This filter is most frequently used for the morphological analysis of comets. It allows the radial and rotational gradients of the sur-face brightness of the images to be calculated from a simple transformation of the coordinates. After processing with this filter, the images lose their photometric information, but the weak structures existing in the cometary coma (e.g., jets and envelopes) can be distinguished. In our study, this filter was used in the following way. The image produced by applying the unsharp mask and the Gaussian blur was processed with the Larson–Sekanina filter with a radius of 0.5 pixel and an angle of 10°. To eliminate the false details from the images prepared for interpretation, each of the filters was separately applied to each of the images.

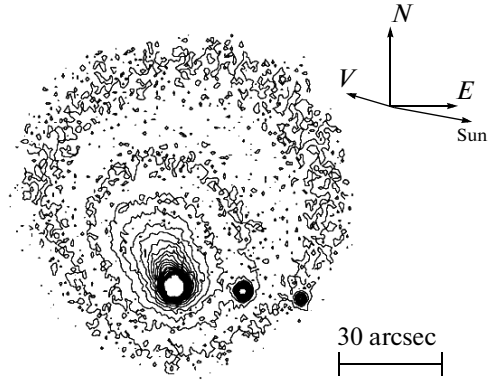

**Fig. 1.** The comet SW1 isophots. The directions to the North, to the East, and to the Sun and the moving direction of the comet ($V$) are indicated. The image of the comet was obtained in the $R$ filter.

Such a technique of selecting the structures in the coma was successfully used by Manzini et al. (2007) for comet C/2002 C1 Ikeya–Zhang and by Korsun et al. (2008; 2010) for comets Schwassmann–Wachmann and C/2003 WT42 (LINEAR).

Figure 2 shows that the application of the digital filters to the images of comet SW1 obtained in December 2008, allows two jets to be distinguished, and these jets are observed during the whole period of record. The jets are directed from the photometric centre of the comet toward the Sun. In the images of the comet acquired in February 2009 (Fig. 3), processing with the digital filters allowed us to select three structures that are directed from the photometric center of the comet away from the Sun.

The digital filters allowed the jets to be reliably distinguished against the background of the bright, structureless coma. However, since these filters are inherently nonlinear, there may be a problem of both the selection of false structures and some spatial shifting of the positions of real jets.

## EVALUATION OF THE ROTATION PERIOD OF THE NUCLEUS

To analyze the angular shifts of the weak-contrast structures in the cometary coma (which allows us to estimate the rotation period of the comet), the cross-correlation method was used. The images were transformed from rectangular to polar coordinates with a centre corresponding to the photometric center of the comet. The radial distance from the nucleus and the azimuth angle counted off anticlockwise from the northward direction were taken as the polar distance and the polar angle, respectively. Moreover, before the coordinate system transformation, all of the images were oriented in the same way, northward and eastward.

To avoid the probably erroneous results in determining the rotation period of the comet, we removed the low-frequency trend in the images of the comet in polar coordinates that were not subjected to digital filtering. The structures selected in such a way were further used for evaluating the rotation period of the cometary nucleus. It is worth noting that we ignore here the motion of each individual particle and consider the dust jet as a united structure, the rotation of which is rigidly connected with the cometary nucleus.

As the result, we selected two jets and three dust structures in the images obtained in 2008 and 2009, respectively. From the photometric profiles, Trigo-Rodrígues et al. (2010) also found that two active zones were present on the surface of the cometary nucleus during the period from 2008 to 2009.

Figures 4a and 5a show the structures superposed with an accounting for the value of the rotation period determined with the cross-correlation method for each of the observation sets.

Figures 4b and 5b show the general pattern of shifts for each of the selected structures and the approximation curve serving as a basis for determining the rotation period for two sets of observations of the comet.

To confirm the reliability of our estimate of the rotation period, it was necessary to check whether the structures selected in the cometary coma remain the same during the whole observation period for each of the sets. In our case, there are two sets: in December 2008, and February 2009. To check this assumption, we used the synchronous detection method that contains the following procedure. If there are clearly seen features in the images, their position angle $PA_i$ is determined for the time moments $T_i$. If the comet rotates with an angular velocity $1/P$ (where $P$ is the determined period), we check whether the precalculated position angle $PA_i = (\Phi_0 + T_i/P)/(2\pi)$ (where $\Phi_0$ is the initial phase) agrees with the observed one.

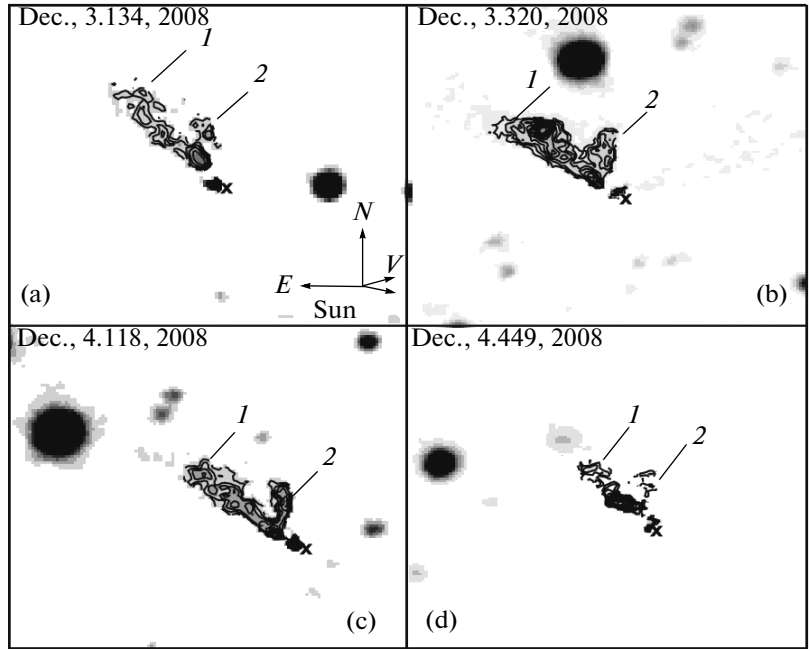

**Fig. 2.** Jets *1* and *2* in the dust coma of comet SW1 selected after the processing of the resulting images of the comet obtained in December 2008. The photometric center of the comet is marked with a cross.

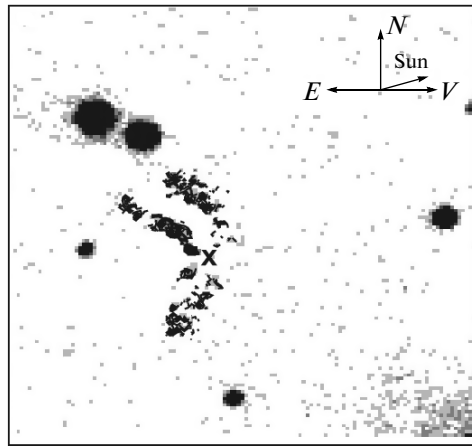

**Fig. 3.** The surface contour of the structures selected in the dust coma of comet SW1 obtained with the digital filter processing of the resulting image of the comet of February 14, 2009. The directions to the North, to the East, and to the Sun and the direction of the comet (*V*) are indicated. The photometric center of the comet is marked with a cross.

As the results of the detection showed, we may suppose that the structures in the images acquired at the 2-meter telescope on February 14 correspond to the structures observed on February 18 and 19. The analogous results were obtained for the observations of December 2008: the structures selected in all of the images of this set can be considered to be the same.

The period estimated for the first (December 3–5, 2008) and second (February 14–19, 2009) sets of observations was $12.1 \pm 1.2$ and $11.7 \pm 1.5$ days, respectively. These estimates agree with each other, and a small scatter in the results is within the limits of measurement error. Such similar estimates obtained for different observation periods demonstrate that the method is rather accurate and provides reliable results for tasks of this kind.

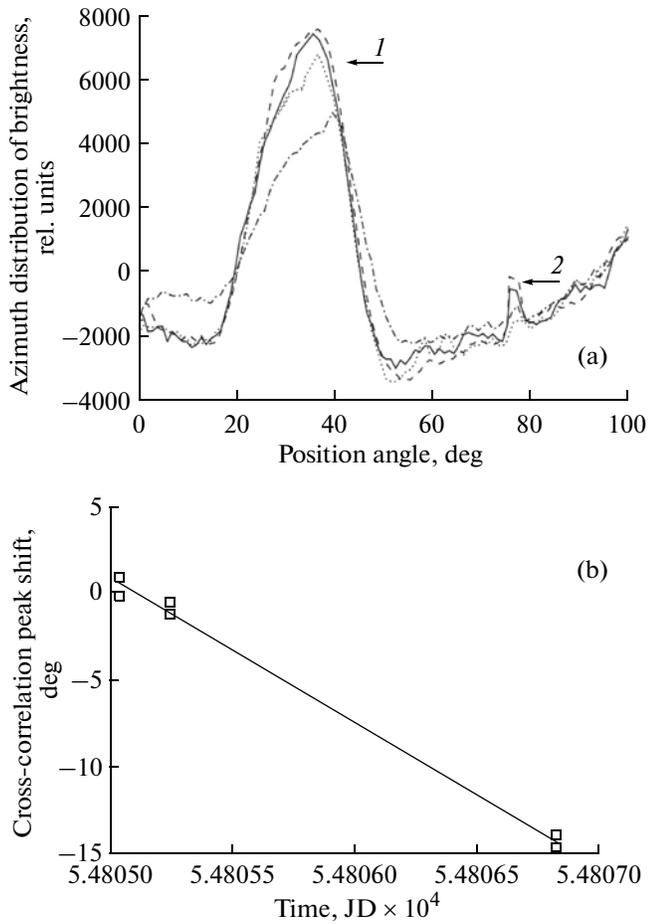

**Fig. 4.** The weak structures in the coma of comet SW1 (the observations taken (a) for December 2008, (b) for February 2009), which were obtained by subtracting from the original images (presented in a polar coordinate system), the smoothed radial profile of these images.

## DISCUSSION AND CONCLUSIONS

To estimate the rotation period of comets is important for understanding and describing the physical processes responsible for the activity of the cometary nucleus. Up to now, this problem also remains actual for the active comet Schwassmann–Wachmann. Many of the researchers attempted to solve this problem by different currently available methods. However, the literature sources present the widely scattered estimates of this parameter. For example, Whipple (1980), Jewitt (1990), and Meech et al. (1993) give short periods for this comet, from 6 to 14.97 hours. In their turn, Moreno (2009) and Trigo-Rodríguez et al. (2010) estimate the period from 50 to 60 days. Trigo-Rodríguez et al. (2010) note that such a scatter in the estimates of the rotation period is most likely con-nected with insufficient or preconceived data sampling used by the previous authors in their analysis. Consequently, for their estimate, Trigo-Rodríguez et al. chose the data obtained in 2008–2010, when the inaccuracy in observations was less than $0.05^m$; and, in addition, they used more precise methods of data processing for selecting the period. Most of the above-described estimates of the rotation period are based on the analysis of the brightness curves of the comet measured in wide filters. Along with these studies, for the comets demonstrating different features in the coma during their active period, the method based on the investigation of the dynamics of the active structures is used. Different variations of this method were earlier applied to the analogous task for comets 29P/Schwassmann–Wachmann 1 (Stansberry et al., 2004), C/1995 O1 Hale–Bopp (Lisse et al., 1999; Samarasinha et al., 2004), P/Swift–Tattle 1992t (Yoshida et al., 1993), etc. In our study, we used one of the versions of this method differing from those mentioned above. Our method for determining the rotation period of comet SW1, as opposed to the other methods, is the most independent of the external parameters used in the other methods, since it is based on the estimate of the shift of weak-contrast structures in the images. For example, Stansberry et al. (2004) apply the method operating with the value of the gas velocity, the variations of which strongly influence the estimate of the rotation period of the comet. The method used by Meech et al. (1993) for evaluating the rotation period of the nucleus is based on a set of estimates of the comet brightness obtained for the whole observation period; however, this set can be influenced by the errors in determining the magnitude of the comet. The only thing, which is required to be taken into account in our method, is strong verification of the fact that the angular shift of precisely the same structures is measured during the whole observation period. This was performed exactly in our study for all of the acquired images of the comet.

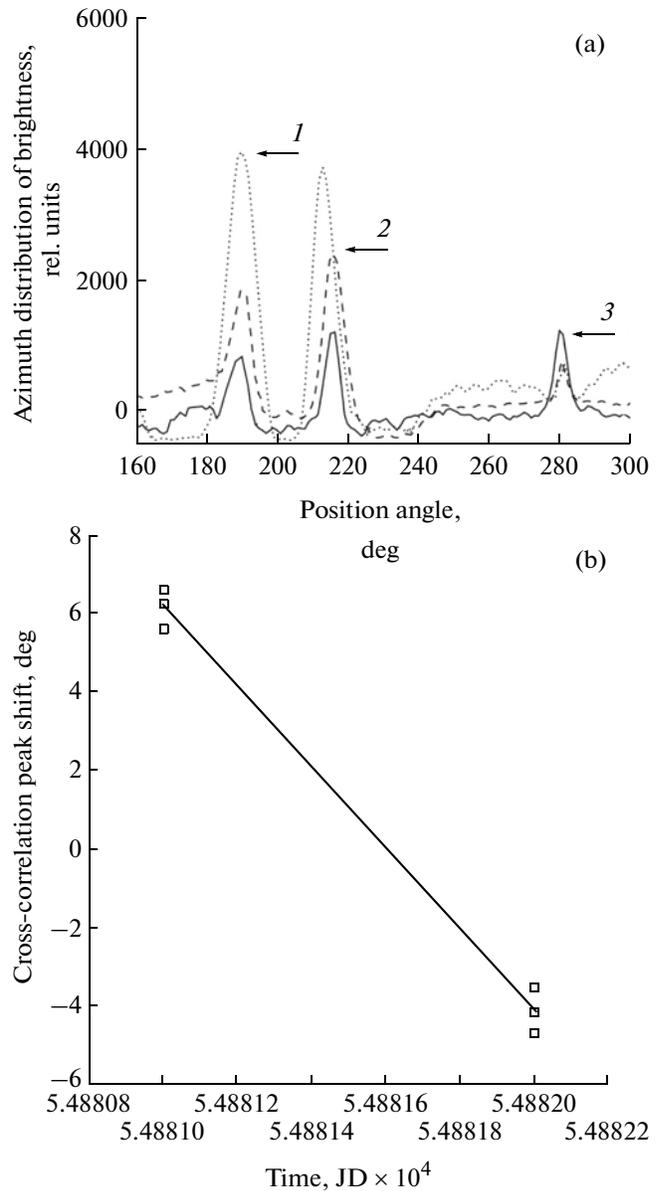

Fig. 5. Displacement of structures in the comet depending on the date of observation and approximation curve to calculate the rotation period of comet SW1, observations which were (a) in December 2008, (b) in February 2009.

We suppose that the present method can be used for evaluating the rotation period of comet SW1 typically demonstrating the long-term flare activity in the form of dust jets. The potential of the use of the dynamics of the active structures in comet SW1 was noticed by Trigo-Rodríguez et al. (2010).

We studied the activity of the comet in 2008 and 2009, and we succeeded in selecting several structures with the use of digital filters and low-frequency trend removal. Two jets and three structures were selected in the images acquired in 2008 and 2009, respectively.

With the cross-correlation method, which allows the shift of the selected weak-contrast features to be estimated, we evaluated the rotation period of the nucleus of comet SW1. According to our results, this period is 12.1 ± 1.2 and 11.7 ± 1.5 days for the observations in 2008 and 2009, respectively. The results for two sets of observations differ only slightly. However, our result differ from those reported by the other authors (Jewitt, 1990; Meech et al., 1993; Luu and Jewitt, 1993; Stansberry et al., 2004; Trigo-Rodríguez et al., 2010). This can be connected with the fact that different authors used different techniques for evaluating the rotation period, and these techniques are influenced by these or other errors.

ACKNOWLEDGMENTS

This work was supported by a grant of the President of Ukraine for young scientists.